\documentclass[useAMS]{mn2e}
\usepackage{graphicx}
\usepackage{latexsym}
\usepackage{url}
\usepackage{psfig}

\title{X-ray ionization rates in protoplanetary discs}

\author[Barbara Ercolano, Alfred E Glassgold]{Barbara Ercolano$^{1,2}$\thanks{E-mail: ercolano@usm.lmu.de (BE)}, Alfred E Glassgold$^{3}$\\
$^{1}$Universit\"ats-Sternwarte M\"unchen, Scheinerstrasse 1, D-81679 M\"unchen\
, Germany\\
$^{2}$xcellence Cluster Universe, Boltzmannstr. 2, D-85748 Garching, Germany\\
$^3$Astronomy Department, University of California, Berkeley, CA 94720}

\begin{document}

\pagerange{\pageref{firstpage}--\pageref{lastpage}} \pubyear{2011}

\maketitle

\label{firstpage}

\def\mnras{MNRAS}
\def\apj{ApJ}
\def\aap{A\&A}
\def\apjl{ApJL}
\def\apjs{ApJS}
\def\bain{BAIN}
\def\araa{ARA\&A}
\def\pasp{PASP}
\def\aj{AJ}
\def\pasj{PASJ}
\def\ga{\sim}

\newcommand{\Lx}{\ensuremath{L_{\rm X}}}
\newcommand{\Tx}{\ensuremath{T_{\rm X}}}
\newcommand{\Nperp}{\ensuremath{N_{\perp}}}
\newcommand{\nh}{n_{\rm H}}
\newcommand{\Nh}{N_{\rm H}}
\newcommand{\Lsun}{\ensuremath{\,L_{\odot}}}
\newcommand{\Msun}{\ensuremath{\,M_{\odot}}}
\newcommand{\Rsun}{\ensuremath{\,R_{\odot}}}
\newcommand{\erg}{\ensuremath{{\rm erg}}}
\newcommand{\ergps}{\ensuremath{{\rm erg}\,{\rm s}^{-1}}}
\newcommand{\sqcm}{\ensuremath{{\rm \,cm}^2}}
\newcommand{\psqcm}{\ensuremath{{\rm cm}^{-2}}}
\newcommand{\ps}{\ensuremath{{\rm s}^{-1}}}
\newcommand{\gpcc}{\ensuremath{\rm \,g\,cm^{-3}}}
\newcommand{\pcc}{\ensuremath{{\rm \,cm}^{-3}}}
\newcommand{\pyr}{\ensuremath{{\rm yr}^{-1}}}
\newcommand{\be}{\begin{equation}}
\newcommand{\ee}{\end{equation}}

\begin{abstract}

Low-mass young stellar objects are powerful emitters of X-rays that can ionize and heat 
the disks and the young planets they harbour. The X-rays produce molecular ions that 
affect the chemistry of the disk atmospheres and their spectroscopic signatures. Deeper 
down, X-rays are the main ionization source and influence the operation of the  
magnetorotational instability, believed to be the main driver for the angular momentum 
redistribution crucial for the accretion and formation of these pre main-sequence stars. X-ray ionization also affects the character of the dead zones around the disk midplane where terrestrial planets are likely to form. To obtain the physical and chemical effects of the stellar X-rays, their propagation through the disk has to be calculated taking into account both absorption and scattering. To date the only calculation of this type was done almost 15 years ago, and  here we present new three-dimensional radiative transfer calculations of X-ray ionization rates in protoplanetary discs. Our study confirms the results from previous work for the same physical parameters. It also updates them by including a more detailed treatment of the radiative transfer and by using ionizing spectra and elemental abundance more appropriate for what is currently known about protoplanetary disks and their host stars. The new calculations for a typical ionizing spectrum yield respectively lower and higher ionisation rates at high and low column densities at a given radius in a disc. The differences can be up to an order of magnitude near 1\,AU, depending on the abundances
used. 



\end{abstract}


\section{Introduction}
Young Stellar Objects (YSOs) are known to be strong emitters of X-rays with typical 
photon energies $\sim$ keV (Preibisch et al. 2005). These X-rays irradiate the accretion disks of YSOs 
where they are responsible for ionizing the inner disk at depths beyond the range of
far ultraviolet (FUV) photons (Glassgold, Najita \& Igea 1997). Using a Monte Carlo 
code, Igea and Glassgold (1999; IG99) studied the propagation of the X-rays in the accretion disks around YSOs (henceforth protoplanetary disks). Including both absorption and scattering, they showed that scattering allows hard X-rays to penetrate disks to depths as far down as vertical columns of $\sim 3 \times 10^{25} \psqcm$, or about 70 g $\psqcm$. This depth extends well below the warm inner atmospheres of protoplanetary disks that are observable at near and mid infrared wavelengths. 

Somewhat surprisingly, there have been no confirming calculations of
the results of IG99 in the intervening years. We do this here using up to date methods embodied in the 
Mocassin code (Ercolano et al.~2003, 2005, 2008a; www.3d-mocassin.net). Ercolano et al.~(2008b, 
2009) and Ercolano \& Owen (2010) have applied this code to a study of diagnostics for protoplanetary disk atmospheres. Mohanty, Ercolano \& Turner (2013) have used the {\sc mocassin} 
X-ray ionization rates in their analysis of midplane ionization of
protoplanetary disks. In addition to verifying and clarifying the results in IG99, we extend 
the scope of this work by studying the dependence on the elemental abundances in protoplanetary disks and by investigating realistic X-ray spectra based on observations of YSOs. We also present our results online in useful form.

\section{Methods}

Our initial objective is to compare results obtained by IG99 with those from 
{\sc moccasin}. It is important to recognize that detailed comparisons are not simple because the two procedures are very different. {\sc mocassin} is a broad based radiative transfer program that includes X-rays, whereas the IG99 calculation was a one-time specific calculation. The underlying physics is treated differently in the two programs and to make matters more difficult, the full electronic version of the IG99 code has been lost.

We performed the ionization rate calculation using a modified version of the 3D radiative transfer and photoionization code {\sc mocassin} (Ercolano et al 2003,
2005, 2008a). The code uses a Monte Carlo photon packet approach to the transfer of radiation (Lucy et al 1999), allowing the treatment of both the 
direct stellar radiation as well as the diffuse fields and the transfer through a mixture of gas and dust. {\sc mocassin} includes 
all the relevant X-ray photoionization and related processes, as
described by Ercolano et al (2008a). This code differs from the
treatment in IG99 by allowing a full treatment of Compton scattering
and secondary ionization by suprathermal electrons that are produced by inner
shell ionization of abundant metals. 
. It also uses a more sophisticated Monte 
Carlo estimator for the radiation field. 

The treatment of physical processes by IG99 is significantly simpler. 
The direct absorption by X-rays is expressed with an analytic approximation 
to the cosmic average cross section for both solar and depleted abundances 
(defined in Table 1). Compton scattering and ionization are also included. The  
secondary electrons that arise from X-ray absorption and the Auger effect 
actually generate the bulk of the ionization, but they are not followed in detail.
Instead the ionization is computed using the semi-empirical energy to make an 
ion-pair, which is close to 36\,eV (Dalgarno et al.~2009) for a fully
neutral gas, which is the case assumed in the models.

In order to compare our calculations with those of IG99 for 
the same or at least very similar conditions, we ran several calculations for monochromatic and thermal spectra using the IG99 depleted abundances (Table 1). We have found that the IG99 analytic approximation for the absorption 
cross-section agrees very well with the {\sc mocassin} cross-section for the same abundances. However, IG99 adopted a low energy cut-off of 1\,keV in their work,  so that the low-energy absorption is different in the two calculations.

Table 1 lists the abundances used in this work. The first column gives the rather extreme abundances used by IG99, where most of the heavy elements are completely 
removed except for obviously highly volatile elements. The scenario behind this 
choice is that the protoplanetary grains have agglomerated and settled to the 
midplane. The third column lists the well determined interstellar abundances for diffuse interstellar clouds obtained from UV absorption line spectroscopy (Savage 
\& Sembach 1996). 

\begin{center}
\begin{tabular}{cccc}    
\multicolumn{4}{c}{Table 1. Abundances} \\
\hline
\hline
Element		& IG99	Depleted$^1$		&   Solar$^2$		&  ISM Depleted$^2$  		\\
\hline	
H	& 1.0				&   1.0				& 1.0				\\
He	& $9.09 \times 10^{-2}$	& $1.00 \times 10^{-1}$	& $1.00 \times 10^{-1}$	\\
C	& $1.49 \times 10^{-4}$	& $2.84 \times 10^{-4}$	& $1.84 \times 10^{-4}$	\\
N	& $3.04 \times 10^{-5}$	& $6.60 \times 10^{-5}$	& $6.00 \times 10^{-5}$	\\
O	& $7.41 \times 10^{-4}$	& $4.90 \times 10^{-4}$	& $3.50 \times 10^{-4}$ \\
Ne	& $1.48 \times 10^{-4}$	& $8.30 \times 10^{-5}$	& $6.90 \times 10^{-5}$ \\
Na	& 0.0				& $1.70 \times 10^{-6}$	& $2.31 \times 10^{-7}$ \\
Mg	& 0.0				& $4.00 \times 10^{-5}$	& $1.00 \times 10^{-6}$ \\
K	& 0.0				& $1.00 \times 10^{-7}$	& $8.57 \times 10^{-9}$ \\
Al	& 0.0				& 				& 			   \\
Si	& 0.0				& $3.20 \times 10^{-5}$	& $1.68 \times 10^{-6}$ \\
S	& 0.0 				& $1.40 \times 10^{-5}$	& $1.40 \times 10^{-5}$ \\
Cl	& 0.0	&	& \\
Ar	& $3.80 \times 10^{-6}$ 	& $2.60 \times 10^{-6}$	& $1.51 \times 10^{-6}$ \\
Ca	& 0.0	&	& \\		
Cr	& 0.0	&	& \\
Fe	& 0.0				& $3.00 \times 10^{-5}$ 	& $1.75 \times 10^{-7}$\\
Ni	& 0.0	&	& \\
\hline
\multicolumn{4}{l}{\small$^1$ Igea \& Glassgold (1999)}\\
\multicolumn{4}{l}{\small$^2$ Savage \& Sembach (1996)}
\end{tabular}
\end{center}

In order to ensure good sampling in the high energy tail of the
spectrum, IG99 opted to perform a set of 20 monochromatic calculations
spanning the range from 1 to 20 keV and combine them by
weighting the resulting ionization rates according to the desired
input spectrum (Green's function). We take a
different approach here by weighting the input photon packets so that all
ranges in the input spectrum are appropriately sampled. This is
important to ensure the accuracy of the ionization rates at large
columns in the disc where only the most energetic photons in the
high-frequency tail of the input spectrum can penetrate.\\

We follow IG99 and assume a ring source of radius 5$R_{\odot}$ placed
at a height of 5$R_{\odot}$. The X-ray luminosity
of a given source is therefore scaled by a factor of two, to account for
two rings placed symmetrically above and below the disc midplane. 
Thus the luminosities quoted in the text and tables are the total for both ring sources. We 
note that the ionization rates do not depend strongly on the exact
choice of the ring radius for values varying between 2 and 10
$R_{\odot}$, as already demonstrated by IG99.\\

The disc density distribution adopted here is the same as in IG99,
i.e. a minimum mass solar nebula model, as in Hayashi, Nakazawa \&
Nakagawa (1985). We assume the same parameters as in IG99 and refer
the reader to Section 2 in their work for a detailed description. We
note here however, that, as demonstrated in IG99, the ionization rates
are not sensitive to the exact choice of the surface density
distribution (see Figure 5 in IG99), and depend mainly on vertical 
column density.  \\

\section{Ionization Rates}

In this section we first make some comparisons between the present
calculations of ionization rates and those in IG99. These involve both 
monochromatic X-rays and those emitted by isothermal sources. Then we will 
consider more realistic spectra based on measurements by the X-ray observatory 
{\it Chandra}, in particular The {\it Chandra} Orion Ultradeep Project 
(COUP; Getman et al.~2005). The parameters of the spectra are summarized in Table 2. 

\begin{center}
\begin{tabular}{cccc}    
\multicolumn{4}{c}{Table 2. X-ray Spectra Parameters} \\
\hline
\hline
Case		& $\Lx (\ergps)$   	& $kT_1$\,( keV) 	&
$kT_2$\,( keV) 	\\
\hline	
Comparison	& $10^{29}$		& 0		& 3, 5, 8			\\
COUP MSM		& $2\times 10^{30}$	& 0.75			& 2.5			\\
COUP Flare		& $2\times 10^{31}$	& 0			& 12			\\
\hline
\end{tabular}
\end{center}

\subsection{Comparison with IG99}

As the basis for their Green's function approach, IG99 carried out 20  
calculations for monochromatic X-rays in the range 1-20\,keV. In Figure 1, 
we compare three of these with the {\sc mocassin} calculations for photon 
energies of 1, 5 and 20 keV. The black {\sc mocassin} points are generally 
in good agreement with the red curves from IG99. The most significant difference 
occurs for 1\,keV, where the IG99 curve dips down too rapidly near 
$N = 10^{22} \psqcm$.

\begin{figure}
\begin{center}
\includegraphics[width=0.47\textwidth]{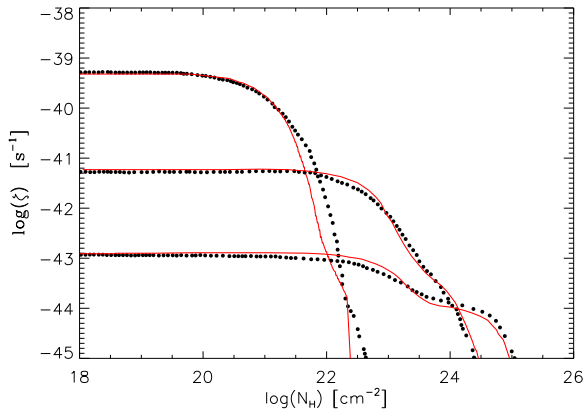}
\caption{Ionization rates plotted vs. vertical column density at 1\,AU for monochromatic beams at 1, 5 and 20 keV. 
Red lines, IG99: black points, this work; both using IG99 depleted abundances (Table 1).}
\label{}
\end{center}
\end{figure}

In the spirit of IG99 Figure 3, we have also calculated the ionization rates for thermal spectra with $kT_X = $ 3, 5 and 8\,keV at three radii, $R =$ 1, 5 and 10\,AU. In Figure 2, 
the calculations are color coded as black (3 keV), red (5 kev) and green (8 keV). The 
10\, AU curve has been shifted down by 1 dex for purposes of clarity. The solid blue curve 
is a result from IG99 for 5\,keV; it was calculated with an assumed
lower energy cutoff of 1\,keV. As noted by IG99, the curves for
different energies do not differ very much, but the IG99 5\,keV curve
is significantly higher than the new calculations at moderately high
densities. Part of the differences can be ascribed to the Green
function approach employed by IG99. As already noted by IG99, their
use of equally spaced $\delta$ functions leads to an undersampling in
the 1-2 keV region when the thermal average is calculated. 
However, taking into account the limitations and uncertainties in making such comparisons, the 
agreement between IG99 and the present calculations is quite good. Figure 2 confirms the 
conclusion of IG99 that the ionization rate depends only weakly on the X-ray temperature and mainly on the X-ray flux, i.e., on the luminosity and the radius.

\begin{figure}
\begin{center}
\includegraphics[width=0.47\textwidth]{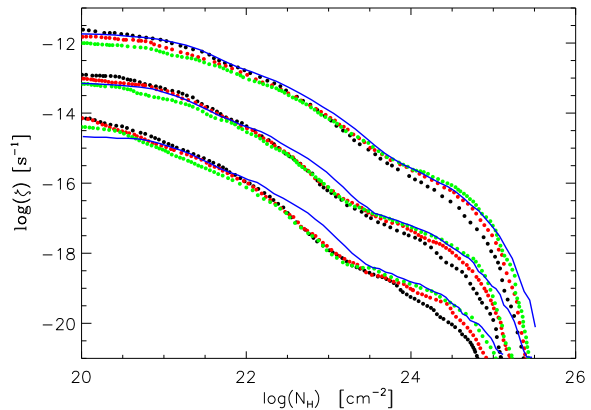}
\caption{ Ionization rates plotted vs. vertical column density for isothermal spectra at 3 keV (black), 5
  keV (red) and 8 keV (green) for 1\,AU, 5\,AU and 10\,AU. The results
  of IG99 for an isothermal spectrum at 5 keV are shown for comparison
  with solid blue lines. For ease of viewing, the 10\,AU results have been shifted down by 10 relative 
to those at 5\,AU.  The X-ray luminosity is $10^{29} \ergps$, consistent with a similar plot in IG99.
}
\label{}
\end{center}
\end{figure}

\subsection{Ionization Rates for COUP Spectra}

The COUP project measured the X-ray properties of 1400 young stars in the Orion 
Nebula Cluster (Getman et al.~2005).  From these data Wolk et al.~(2005) made a 
detailed study of 28 Sun-like stars, and we use the average X-ray properties of 
this sample (their Table 4) for the present calculations. In Table 2 we refer to 
this average spectrum as COUP MSM (for mean solar mass). The low and high X-ray 
temperature components have, respectively, 1/3 and 2/3 of the total luminosity of 
$\Lx = 2 \times 10^{30} \ergps $.
The results are shown in Figure 3 for solar abundances (black curves) and ISM 
depleted abundances (red curves). At small and moderately large column densities, 
$\log \Nh < 23.5 \psqcm$, there is a fairly strong dependence on abundance. Recalling the 
discussion in Sec.~2 on grain growth and settling, this may also be considered as a dependence on grain size.
Solar abundances might then apply to {\it small} grains where the heavy elements 
are fully exposed to the X-rays independent of their partitioning between gas and 
grain. The depleted abundances would correspond to the situation where the depleted 
elements are incorporated into grains that grow and settle out close to the midplane. 
This is probably the case for protoplanetary disk atmospheres exposed to stellar X-rays.
For very large column densities the ionization rates are almost independent of the
abundances, as might be expected for the situation where the main interaction of
the X-rays is Thomson/Compton scattering and ionization of H and He. Figure 3 confirms 
again that the Thomson cross section determines the range of the X-rays, in this 
$\log \Nperp \sim 25.5 \psqcm$. This is expected due to the significant hardening of the radiation field at these columns. Thomson scattering begins to dominate over absorption for energies greater than $\sim$5keV, but photons of these energies are the only ones that can penetrate so deep.
 Some of the curves in Figure 3 seem to drop precipitously due to 
the small number of surviving photons at the largest columns. At these maximum depths,
the ionization rate has reached $\zeta = 10^{-22}\, \ps$.

\begin{figure}
\begin{center}
\includegraphics[width=0.47\textwidth]{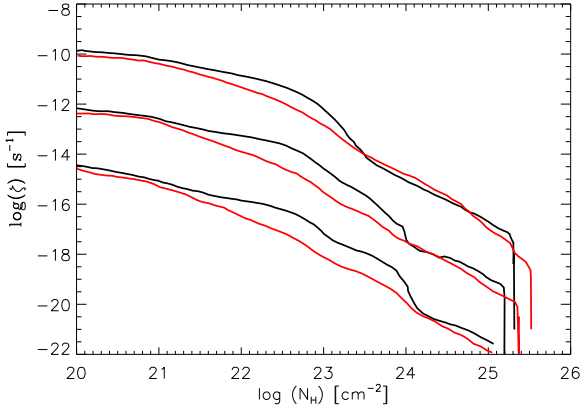}
\caption{Ionization rates plotted vs. vertical column density for a two-temperature average representation of the COUP X-ray observations (Wolk et al.~2005) of the Orion Nebula Custer at 1\,AU, 5\,AU and 10\,AU 
for solar abundances (black) and depleted interstellar abundances (red) given in Table 1. 
The X-ray luminosity is $\Lx = 2 \times 10^{30} \ergps$, and the other spectrum parameters 
are given in Table 2. For ease of viewing, the 5\,AU and 10\, AU curves have been shifted 
down by factors of 10 and 1,000 relative to those for 1\,AU. }
\label{}
\end{center}
\end{figure}

We have also calculated the ionization rate for the average flare spectrum using Table 2 
of Getman et al.~(2008). The spectrum parameters are given in Table 2 (under COUP Flare), 
and the results shown in Figure 4 following the style of the previous figure. The results 
are roughly similar to Figure 3, the non-flaring case of COUP results for solar mass 
YSOs. Due to the high $\Lx$ and high $\Tx$, the results are even more sensitive to 
abundances, and the dominance of Compton/Thomson scattering is not manifest until very 
large vertical columns.  

\begin{figure}
\begin{center}
\includegraphics[width=0.47\textwidth]{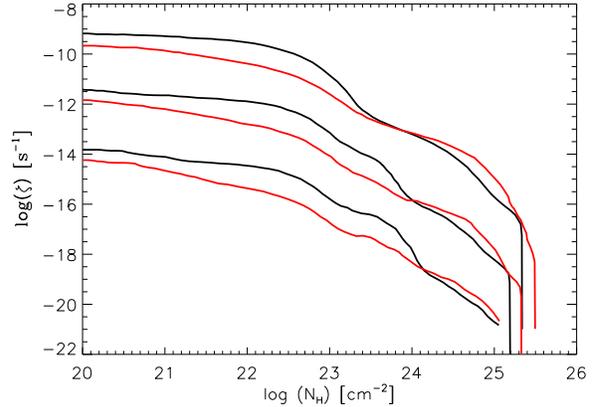}
\caption{Ionization rates plotted vs. vertical column density for a typical one-temperature flare spectrum (Getman et al.~2008) 
at 1\,AU, 5\,AU and 10\,AU for solar abundances (black) and depleted ISM abundances (red) (Table 1). The X-ray luminosity is $\Lx = 2 \times 10^{31} \ergps$, and the X-ray temperature is given in Table 2. For ease of viewing, the 5\,AU and 10\, AU curves have been shifted down by factors of 10 and 1,000 relative to those 
for 1\,AU.}
\label{}
\end{center}
\end{figure}

The influence of the ionizing spectrum on the ionisation rates in the disc can be better appreciated by comparing the results for a typical one-temperature flare spectrum (Getman et al.~2008) against those for a two-temperature representation of the average spectrum of the COUP 
solar-mass sample (Wolk et al.~2005) for the Orion Nebula Custer. This comparison is illustrated in Figure 5 at 1\,AU, 5\,AU and 10\,AU using depleted abundances. 
The luminosities were normalised to $1 \,\ergps$ ionising annulus (i.e., a total luminosity of 
$2 \, \ergps$). The harder flare spectrum produces higher ionisation rates at larger columns. 
The radiative transfer is dominated by Thompson/Compton scattering in this case, as is clearly visible in the bump in the ionisation rate curves at high column density. The larger (by at most
1 dex) ionisation rates for the flare case in the range $\log \Nh \sim 23-25 \, \psqcm$ 
are compensated by smaller ionisation rates for the same case at low column densities, 
$\log \Nh \sim 21 \, \psqcm$ for the solar abundances (not shown) and 
$\log \Nh \sim 22 \, \psqcm$ for depleted abundances (Figure 5). The differences are due to the fact that the harder photons require a larger column before they can be efficiently absorbed by the gas. The transition occurs at lower columns in the solar case compared to the depleted abundance case, i.e. $\log \Nh \sim 21 \, \psqcm$ compared to $\log \Nh \sim 22 \, \psqcm$,
because the higher metal abundances allows the required optical depth for interaction to be achieved at lower hydrogen columns.

\begin{figure}
\begin{center}
\includegraphics[width=0.47\textwidth]{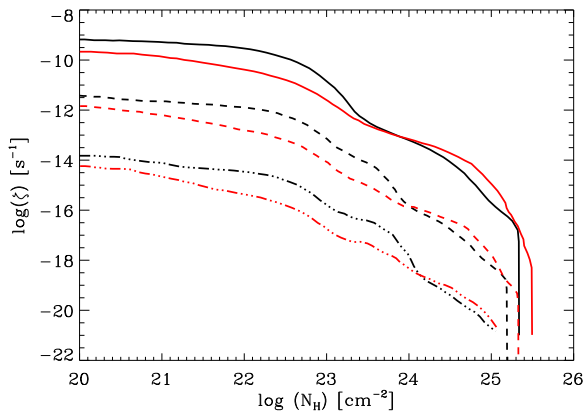}
\caption{Ionization rates plotted vs. vertical column density for a typical one-temperature flare spectrum (Getman et al.~2008) at 1\,AU, 5\,AU and 10\,AU (black lines) and for a two-temperature average representation of the COUP X-ray observations (Wolk et al.~2005) of the Orion Nebula Custer (red lines). All models employed depleted ISM abundances (Table 1). The X-ray luminosity is $\Lx = 2\, \ergps$ for both cases, and the X-ray temperatures are given in Table 2. For ease of viewing, the 5\,AU and 10\, AU curves have been shifted down by factors of 10 and 1,000 relative to those for 1\,AU.}
\label{}
\end{center}
\end{figure}

\section{Conclusions}

\begin{figure}
\begin{center}
\includegraphics[width=0.47\textwidth]{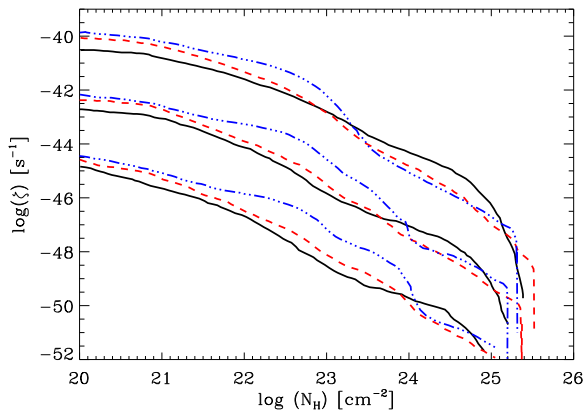}
\caption{Ionization rates plotted vs. vertical column density for a 5
  keV thermal spectrum with IG99 depleted abundances
  1\,AU, 5\,AU and 10\,AU (black lines) compared to a two-temperature
  average representation of the COUP X-ray observations (Wolk et
  al.~2005) of the Orion Nebula Custer for depleted ISM abundances
  (red dashed lines) and solar abundances (blue dash-double-dot
  lines). Abundances are listed in Table 1. The X-ray luminosity is $\Lx = 2\, \ergps$ in all cases, and the X-ray temperatures are given in Table 2. For ease of viewing, the 5\,AU and 10\, AU curves have been shifted down by factors of 10 and 1,000 relative to those for 1\,AU.}
\label{}
\end{center}
\end{figure}

We have performed new three-dimensional radiative transfer
calculations of X-ray ionization rates in protoplanetary discs. Our
results are in broad agreement with the previous calculations of IG99
for the same stellar and disc properties. 

The calculations include a two-temperature average representation of the 
COUP X-ray sources in the Orion Nebula Custer (Wolk et al.~2005). In 
essentially all cases, the effects of the X-rays fade by a vertical depth 
$\sim 70$\, gr cm$^{-2}$ at 1 AU, a very small fraction of the disc thickness 
at that radius. This finite depth occurs because penetrating high-energy X-rays interact 
with matter through the Compton effect, rather than by photoelectric absorption.   
The ionization rate at such depths for a typical YSO is $\sim 10^{-22} \ps$,
significantly smaller than the potential contributions from the radiative decay 
of galactic $^{26}$Al (Stepinski 1992) or shielded cosmic rays (Cleeves et al.~2013). 
As shown in the figures, the effects of scattering appear at even smaller vertical 
columns near $10^{24} \psqcm$, consistent with the Thomson cross section.

There are significant differences in the ionisation rates between the simple 
one-temperature thermal spectra used by IG99 and the two-temperature fits to 
the X-ray observations. In particular, after normalising to the same X-ray 
luminosity, the ionisation rates based on the COUP observtaions are systematically 
higher at large column densities and systematically lower at low column 
densities than the IG99 calculations (see Figure~6). The change occurs at typical column 
densities that depend on the radius and on the abundances used. Typical 
differences are of order half a magnitude for calculations employing depleted 
ISM abundances (Table 1) and a little more than a magnitude for solar
abundances, as shown in Figure~6. The effects on the electron abundance will be less, however,  
because it depends on the square root of the ionization rate 
(e.g.~\'Ad\'amkovics et al.~2012).

The calculations in this paper are relevant for a number of astrophysical problems.
In the upper atmosphere of the inner disc ($ N < 10^{24} \psqcm$), X-ray ionization affects the chemistry 
and thus the molecular spectroscopy of the disc. Below, the X-rays help determine the 
geometry of the dead-zone associated with the magneto-rotational instability, especially 
its vertical thickness. On the other hand, stellar X-rays do not play a role on the 
ionization of the mid-plane region of the dead-zone, where other processes may be at 
work (e.g., Turner \& Drake 2009; Mohanty et al.~2013). Another interesting application 
of the ionization by X-rays generated by young stars is the atmospheres of gaseous 
planets formed in protoplanetary discs, including the young Sun (e.g., Turner et al.~2013). 

In order to make the results in this paper available for such applications, they 
are posted as tables in the online material associated with this
article, as well as at www.3d-mocassin.net. New calculations with
different ionisation and disc or planetary atmosphere properties will be readily performed upon
request.

\section{Acknowledgements}

This work has been supported in part by NASA grant NNG06GF88G (Origins) to the University 
of California. 


\newpage

\begin{thebibliography}{}

\bibitem[]{jdkdjel} \'Ad\'amkovics, M., Glassgold, A. E. \& Meijerink, R.
\ 2011, ApJ, 736:143
\bibitem[Cleeves et al.(2013)]{2013ApJ...772....5C} Cleeves, L.~I., Adams, 
F.~C., \& Bergin, E.~A.\ 2013, \apj, 772, 5 
\bibitem[] {hdjkwhe} Dalgarno, A., Yan, M. \& Liu, W.-H. 1999, ApJS, 125, 237	
\bibitem[Ercolano et al.(2003)]{2003MNRAS.340.1136E} Ercolano, B., Barlow, 
M.~J., Storey, P.~J., \& Liu, X.-W.\ 2003, \mnras, 340, 1136 
\bibitem[Ercolano et al.(2005)]{2005MNRAS.362.1038E} Ercolano, B., Barlow, 
M.~J., \& Storey, P.~J.\ 2005, \mnras, 362, 1038 
\bibitem[Ercolano et al.(2008)]{2008ApJS..175..534E} Ercolano, B., Young, 
P.~R., Drake, J.~J., \& Raymond, J.~C.\ 2008a, \apjs, 175, 534
\bibitem[Ercolano et al.(2008)]{2008ApJ...688..398E} Ercolano, B., Drake, 
J.~J., Raymond, J.~C., \& Clarke, C.~C.\ 2008b, \apj, 688, 398
\bibitem[Ercolano 
\& Owen(2010)]{2010MNRAS.406.1553E} Ercolano, B., \& Owen, J.~E.\ 2010, \mnras, 406, 1553 
\bibitem[] {dfe} Getman, K. et al. 2005, AppJS, 160, 319 
\bibitem[] {edc} Getman K., et al. 2008, ApJ, 688, 437 
\bibitem[Glassgold et al.(1997)]{1997ApJ...480..344G} Glassgold, A.~E., 
Najita, J., \& Igea, J.\ 1997, \apj, 480, 344 
\bibitem[] {ndek} Hayashi, C., Nakazawa, K., \& Nakagawa, Y. 1985, in Protostars \& Planets II, ed. D. C. Black \& M. S. Mathews (Tucson: Univ. Arizona Press), 1100	
\bibitem[] {dhjke} Igea, J. \& Glassgold, A. E. 1999, ApJ, 518, 858 	
\bibitem[] {neaklcd} Lucy L. 1999, A\&A, 345, 211				
\bibitem[] {nedklcd} Mohanty, S., Ercolano, B. \& Turner, N. J. 2013, ApJ, 764:65
(25pp) 
\bibitem[Preibisch et al.(2005)]{2005ApJS..160..401P} Preibisch, T., Kim, 
Y.-C., Favata, F., et al.\ 2005, \apjs, 160, 401 
\bibitem[] {nesklcd}  Savage, B. D. \& Sembach, K. R. 1996, ARAA, 34, 279  
\bibitem[] {hgjh} Stepinski, T. F. Icarus 97, 130,  1992			
\bibitem[Turner 
\& Drake(2009)]{2009ApJ...703.2152T} Turner, N.~J., \& Drake, J.~F.\ 2009, \apj, 703, 2152 
\bibitem[] {neklcd} Turner, N. J., Lee, M. H. \& Sano, T. 2013, arXiv 1306.2276	
\bibitem[] {de} Wolk, S. et al. 2005, ApJ, 160, 423	
\end{thebibliography}

\label{lastpage}

\end{document}